\documentclass[twoside,a4paper,11pt]{sea9}
\usepackage{graphicx}
\usepackage{hyperref}
\usepackage{movie15}  
\begin{document}
\pagenumbering{arabic}
\pagestyle{myheadings}
\thispagestyle{empty}
{\flushleft\includegraphics[width=\textwidth,bb=58 650 590 680]{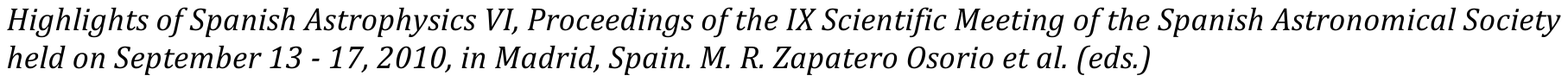}}
\vspace*{0.2cm}
\begin{flushleft}
{\bf {\LARGE
%
Spectroscopic properties of nearby
late-type stars, members of stellar
kinematic groups. 
%
}\\
\vspace*{1cm}
%
J.~Maldonado$^{1}$,
R.M.~ Mart\'inez-Arn\'aiz$^{2}$, 
C.~Eiroa$^{1}$,
D.~Montes$^{2}$,
and 
B.~Montesinos$^{3}$
%
}\\
\vspace*{0.5cm}
%
$^{1}$
Universidad Aut\'onoma de Madrid, Dpto. F\'isica Te\'orica, M\'odulo 15,
Facultad de Ciencias, Campus de Cantoblanco, E-28049 Madrid, Spain,
\textsf{jesus.maldonado@uam.es}\\
$^{2}$
Universidad Complutense de Madrid, Facultad Ciencias F\'isicas,
E-28040 Madrid, Spain \\
$^{3}$
Laboratorio de Astrof\'isica Estelar y Exoplanetas, Centro de Astrobiolog\'ia, LAEX-CAB (CSIC-INTA),
ESAC Campus, P.O. BOX 78, E-28691, Villanueva de la Ca\~{n}ada, Madrid, Spain
%
\end{flushleft}
%
\markboth{
Spectroscopic properties of nearby late-type stars
}{ 
%
J.~Maldonado et al. 
%
}
\thispagestyle{empty}
\vspace*{0.4cm}
\begin{minipage}[l]{0.09\textwidth}
\ 
\end{minipage}
\begin{minipage}[r]{0.9\textwidth}
\vspace{1cm}
\section*{Abstract}{\small
%
 Nearby late-type stars are excellent targets to look for young objects in stellar associations
 and moving groups. The study of these groups goes back more than one century ago
 however, their origin is still misunderstood.
 Although their existence have been confirmed by statistical studies of large sample of stars,
 the identification of a group of stars as member of moving groups, is not an easy task, list
 of members often change with time and most members have been identified by means of
 kinematics criteria which is not sufficient since many old stars can share the same spatial
 motion of those stars in moving groups.
 In this contribution we attempt to identify unambiguous moving groups members,
 among a sample of nearby-late type stars. High resolution echelle spectra
 is used to i) derive accurate radial velocities which allow us to study
 the stars' kinematics and make a first selection of moving groups members;
 and ii) analyze several age-related properties for young late-type stars
 (i.e., lithium  Li~{\sc i} \space 6707.8\AA \space line, $R'_{\rm HK}$ index).
 The different age-estimators are compared and the moving group
 membership of the kinematically selected candidates are discussed.
%
\normalsize}
\end{minipage}
%
%
%

\section{Introduction \label{intro}}
 
 The dispersal of young stars out into the field and whether or not there are kinematically-related
 groups of stars in the solar neighbourhood is one of the open problems in Galactic astrophysics. 
 Although
 stars do not form isolated but within clusters and associations,  most of them end up in the field. 
 Once scattered through the Galaxy, these young stars are difficult to identify. 
 Eggen \cite{eggen94} assumed that clusters form a halo of evaporated
 stars. During its disintegration, the  whole group can not be identified although within
 some regions one can found stars moving in the same direction and at the same rate. 
 This idea gave rise to the concept of moving group or Stellar Kinematic Group (hereafter SKG).\\
 
 The idea of a moving group as a group of stars sharing a common origin has been the subject of
 an intense discussion. It is now, however, well established that the ``classical'' moving groups
 (e.g. Hyades, Ursa Major)
 are in reality a mixture of two different populations: a group of coeval stars
 (related to the halo of an evaporating cluster) and a second group with a dynamical (resonant) origin
 (e.g. Famaey et al. \cite{famaey}).\\

 Identifying stars with a common origin (i.e. same kinematics, same age, same chemical composition)  is only possible if a combination of techniques are used.
 Nearby-late type stars are excellent targets for this kind of study since i)  
 their spectra is 
 full of narrow absorption lines, allowing determination of accurate radial velocities,
 and ii)  it is unlikely that an old star by chance shares chromospheric indices or
 a lithium abundance similar to those of  young solar-like stars.

\section{How can we identify stars in SKGs?} 

  In a recent work, Maldonado et al. \cite{maldonado10} studied a sample of 405 nearby late-type stars,
  attempting to identify unambiguous SKGs members by analysing high-resolution (R$\sim$57000) echelle
  spectra obtained in 2-3 meters class telescopes.
  The study is focused in nearby (distances less than 25 pc), main-sequence (luminosity class V/IV-V),
  late-type (spectral types FGK) stars. The Hipparcos catalogue \cite{hipparcos} is used as a reference. 
  To identify stars in SKGs, Maldonado et al. \cite{maldonado10} analyzed both the kinematics and the spectroscopic age indicators.
       
\subsection{Kinematic analysis} 

   Radial velocities were measured by cross-correlating, using the IRAF routine \textit{fxcor},
   the spectra of the program stars with spectra of radial velocity standard stars of similar
   spectral types. For those known spectroscopic binaries, the radial velocity of the centre of 
   mass of the system were considered. Typical uncertainties are
   between 0.15 and 0.25 $\rm{km s^{-1}}$. These radial velocities were used together with Hipparcos
   parallaxes \cite{Leeuwen} and Tycho-2 proper motions \cite{tycho2} to compute
   the Galactic-spatial velocity components $(U,V,W)$, as explained in Montes et al. \cite{davida}.\\

\begin{figure}[ht!]
\center
\includegraphics[scale=0.47]{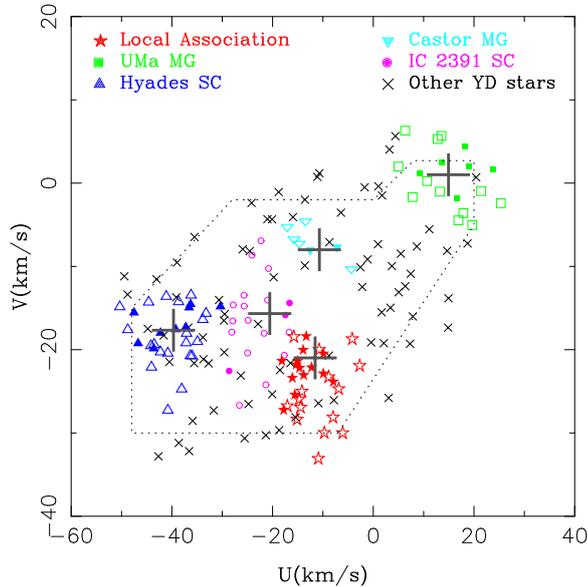}
\caption{\label{fig1} 
$(U,V)$  plane.
Different colours and symbols indicate membership to different SKGs.
Large crosses represent the convergence point of the young SKGs shown in the figure as
given by Montes et al. \cite{davida}.
The dotted line represents the boundary of the young disc population.
Figure from Maldonado et al. \cite{maldonado10}.
}
\end{figure}

   Young stars are assembled in an specific region of the $(U,V)$ plane $(-50 \ \rm{km s^{-1}}\!<\!U\!<\!20 \ \rm{km s^{-1}};
  -30 \ \rm{km s^{-1}}\!<\!V\!<\!0 \ \rm{km s^{-1}})$, although the shape is not a square (see Fig.~\ref{fig1}).
   Possible members of SKGs are selected allowing a dispersion of 8 $\rm{kms^{-1}}$ in the 
    $U$, $V$ components  with respect to the central position
   of the SKG. The same dispersion is considered when taking the $W$ component into account.
   The final number of candidates for each SKG is given in Table~\ref{tab1} (column 4).


\subsection{Age estimates} 

  Members of a given SKG should be coeval and since clusters disperse on time scales
  of a few hundred years they should also be moderately young ($\sim$ 50 - 650 Myr).
  The classic method to compute
  the stellar age (i.e. evolutionary tracks) does not work good enough for the latest spectral types.
  However, late-type stars show other properties which can be used to determine their age:\\  
 
  \textbf{Lithium abundance:}  An  age estimate of late-type stars can be carried out by comparing their
  Li~{\sc i} 6707.8 \AA \space  equivalent width with those of stars in 
  well known young open clusters of different ages (e.g. Maldonado et al., Montes et al. \cite{maldonado10,davidb}).
  Nevertheless, it should be regarded as an additional age indicator since the relation lithium-age is poor constrain and biased
  towards younger ages.

  \textbf{Age derived from cromospheric activity:} There are several observables
  of the magnetic field of a solar-type star: chromospheric emission lines, e.g. 
  Ca~{\sc II} H, \& K or Ca~{\sc II} IRT, or the X-ray emission from the stellar corona
  (e.g. L\'opez-Santiago et al., Mart\'inez-Arn\'aiz et al. \cite{javi,raquel})
  In addition, there is a strong correlation between the stellar rotation and the
  chromospheric activity in cool stars (e.g. Montesinos et al. \cite{benja}). In this way,
  the stellar age can be estimated (Mamajek \& Hillenbrand \cite{mamajek}, Garc\'es et al. \cite{garces}):

  \begin{itemize}
  \item By using the index $R'_{\rm HK}$ which is a measure of the cromospheric emission
        in the cores of the {\rm Ca~{II} H, \& K} absorption lines, normalised to the
        total bolometric emission of the star. 

  \item By searching for X-ray counterparts in the ROSAT catalogue. 

  \item From the rotational period of the star (gyrochronology) 
 
  \end{itemize}

\begin{figure} 
\center
\includegraphics[angle=270,scale=0.47]{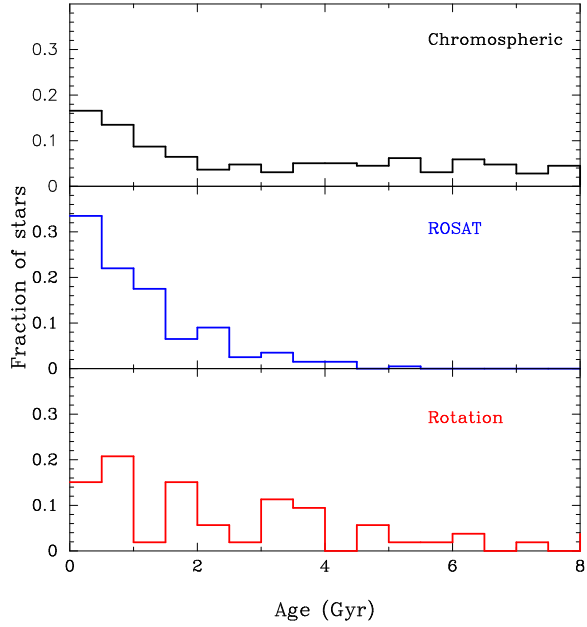}
\caption{\label{fig2} Age distribution for chromospheric-derived ages (upper panel), ROSAT ages
(middle panel), and rotational ages (lower panel).
Figure from Maldonado et al. \cite{maldonado10}.
}
\end{figure}

  The agreement between the different activity-age estimates is overall
  good, as can be seen in Fig.~\ref{fig2}.  
  Chromospheric age shows an enhancement of the star formation rate
  in the last 2 Gyr, then the distribution becomes more or less flat.
  ROSAT ages are biased towards stars younger than 3-4 Gyr;
  i.e., older stars have negligible (or undetectable) X-ray emission, and therefore
  their distribution does not offer information on the stellar formation history.
  As far as rotational ages are concerned, there are not enough stars with measured rotational
  periods to draw robust conclusions.\\

  A summary of the ages obtained (percentages of stars according to their ages for each
  age-indicator) can be seen in Table~\ref{tab2}.


\section{Conclusions and prospects for future work}

   From a total sample of 405 stars, Maldonado et al. \cite{maldonado10} identify 102 stars which share
   the same kinematics that those stars in SKGs (i.e. $\sim$ 25\% of the sample). In addition, 78 stars
   are classified as \textit{other young discs stars} (i.e. stars which are in the boundaries of the young disc
   population but without a clear identification to some of the SKGs).   
   From them, 36 have ages that agree with the accepted ages of the corresponding moving group. That
   means that only $\sim$10\% of the nearby late-type stars can be associated to SKGs. 
   Table~\ref{tab1} summarises the number of kinematic candidates to the different SKGs and the final number of possible
   members and no-members of each of the SKGs studied in Maldonado et al. \cite{maldonado10}.\\ 

   List of young stars in SKGs can be very useful for further investigations. Some examples 
   include search
   for solar analogues, substellar companions, study of the flux-flux and rotation-activity-age
   relationships in groups of stars with different ages  (Mart\'inez-Arn\'aiz et al. \cite{raquel})
   or search for cold faint dusty debris
   discs (e.g. the DUNES project, Eiroa et al., \cite{dunes}, Montesinos et al. \cite{benja10}).


\begin{table} 
\caption{Number of stars identified as possible members and non-members of moderately young SKGs.}          
\center
\center
\begin{tabular}{ccccc}
\hline\hline
Kinematic Group   & Age (Myr)  &  Candidates & Possible Members & No-members  \\ [0.5ex]   
\hline
IC2391            & 35-55      &  19   &  6  &  4 \\
Castor            & 200        &  7    &  4  &  0 \\
Ursa Major        & 300        &  18   &  6  &  6 \\
Local Association & 20-150     &  29   & 14  &  8 \\
Hyades            & 600        &  29   & 11  &  9 \\ [1ex]  
\hline
\end{tabular}
\label{tab1}
\end{table}


\begin{table} 
\caption{Percentages of stars according to their estimated age for each age indicator.}
\center
\center
\begin{tabular}{ccccc}
\hline\hline
Age-indicator     & $\sim$ 80-100 Myr  &  $\sim$ 300 Myr & $\sim$ 665 Myr & $>$ 665 Myr \\ [0.5ex]   
\hline
Lithium$^{\dag}$          &    4       &            8    &                &      23     \\  
Chromospheric             &    5       &            10    &     13        &      72     \\
ROSAT                     &   23       &                 &      51        &      26     \\
Rotation$^{\ddag}$        &   22       &                 &      28        &      50     \\ 
\hline
\multicolumn{5}{l}{$^{\dag}$ \small{50\% older than 80-100 Myr; 15\% of the stars show no photospheric Li~{\sc I}}}\\
\multicolumn{5}{l}{$^{\ddag}$ \small{Rotational periods only available for roughly 17\% of the whole sample}}\\
\hline
\end{tabular}
\label{tab2}
\end{table}

%
%
\small  
%
\section*{Acknowledgments}   
%

  This work were supported by the Spanish Ministerio de Ciencia e Innovaci\'on (MICINN),
  Plan Nacional de Astronom\'ia y Astrof\'isica, under grants \emph{AYA2008-01727} and 
  \emph{AYA2008-00695} and the Comunidad Aut\'onoma de
  Madrid, under PRICIT project \emph{S-2009/ESP-1496} (AstroMadrid).
  J.M. acknowledges support from the Universidad Aut\'onoma de Madrid
  (Department of Theoretical Physics).

%

%

\begin{thebibliography}{}
\small
%

\bibitem{dunes}{Eiroa, C., Fedele., D., Maldonado, J., and the DUNES consortia. 2010, A\&A, 518, L131} 

\bibitem{eggen94}{Eggen, O. J. 1994, in Galactic and Solar System Optical Astrometry, ed.
                  L. V. Morrison \& G. F. Gilmore, 191} 

\bibitem{hipparcos}{ESA, ed. 1997, ESA Special Publication Vol. 1200, The HIPPARCOS and TYCHO catalogues. Astrometric and photometric star catalogues derived from the ESA HIPPARCOS Space Astrometry Mission}

\bibitem{famaey}{Famaey, B., Pont, F., Luri, X., Udry, S., Mayor, M., Jorissen, A. 2007, A\&A, 461, 957}

\bibitem{garces}{Garc\'es, A., et al. 2010, in prep}

\bibitem{tycho2}{H{\o}g, E., Fabricius, C., Makarov, V.~V., Urban, S., Corbin, T., Wycoff, G., Bastian, U., Schwekendiek, P.,
Wicenec, A. 2000, A\&A, 355L, L27}

\bibitem{javi}{L\'opez-Santiago, J., Montes, D., G\'alvez-Ortiz, M. C., et al. 2010, A\&A, 514, A97}

\bibitem{maldonado10}{Maldonado, J., Mart\'inez-Arn\'aiz, R. M., Eiroa, C., Montes, D., Montesinos, B. 2010, A\&A, 521, A12}

\bibitem{mamajek}{Mamajek, E.~E. \& Hillenbrand, L.~A. 2008, ApJ, 687, 1264}

\bibitem{raquel}{Mart\'inez-Arn\'aiz, R. M., Maldonado, J., Montes, D.,  Eiroa, C.,  Montesinos, B. 2010, A\&A, 520, A79}

\bibitem{david10}{Montes, D. 2010, this proceedings}

\bibitem{davida}{Montes, D., L\'opez-Santiago, J., G\'alvez, M. C., et al., 2001a, MNRAS, 328, 45}

\bibitem{davidb}{Montes, D., L\'opez-Santiago, J., Fern\'andez-Figueroa, M. J., \& G\'alvez, M. C. 2001b, A\&A, 379, 976}

\bibitem{benja10}{Montesinos, B., and the DUNES consortia. 2010, this proceedings}

\bibitem{benja}{Montesinos, B., Thomas, J.~H., Ventura, P., Mazzitelli, I. 2001, MNRAS, 326, 877}

\bibitem{Leeuwen}{van Leeuwen, F. 2007, Hipparcos, the New Reduction of the Raw Data, Astrophysics and Space Science Library, Vol. 350,
XXXII, 449 p., Hardcover, ISBN: 978-1-4020-6341-1}


%
%
\end{thebibliography}
\end{document}